\documentstyle[preprint,eqsecnum,aps,epsfig]{revtex}

\tightenlines

\newcommand{\cs}{\'{c}}

\newcommand{\beq}{\begin{equation}}
\newcommand{\eeq}{\end{equation}}
\newcommand{\bdm}{\begin{displaymath}}
\newcommand{\edm}{\end{displaymath}}
\newcommand{\beqa}{\begin{eqnarray}}
\newcommand{\eeqa}{\end{eqnarray}}
\newcommand{\beqab}{\begin{eqnarray*}}
\newcommand{\eeqab}{\end{eqnarray*}}

\def\nn{\nonumber}

 \def\@makefnmark{\hbox to 0pt{$^{\@thefnmark}$\hss}}  

\def\R4{\overline R}

\newcounter{saveeqn}%

\begin{document}

\draft



\preprint{NYU-TH-02-03-11}

\title{DGP BRANE AS A GRAVITY CONDUCTOR}

\author{ 
Marko Kolanovi\cs \footnote{e-mail: mk679@nyu.edu} }
\address{Department of Physics, New York University,
New York, NY 10003}

\date{\today}
\maketitle
\begin{abstract}

We study how the DGP (Dvali-Gabadadze-Porrati) brane affects
particle dynamics in linearized approximation. We find that
once the particle is removed from the brane
it is repelled to the bulk. Assuming that the
cutoff for gravitational interaction
is $M_*\sim 1/\epsilon$, we calculate the classical self energy of
a particle as the function of its position. Since the particle
wants to go to the region where its self energy is lower, it is repelled
from the brane to the bulk where it gains its 5D self energy. Cases when mass 
of the particle $m<8\pi^2M_*$
and $m>8\pi^2M_*$ are qualitatively different, and in later case
one has to take into account effects of strong gravity. In both
cases the particle is repelled from the brane.
For $m<8\pi^2M_*$ we obtain the same result from the 'electrostatic'
analog of the theory. In that language mass (charge) in the bulk
induces  charge distribution on the brane which shields the other
side of the brane and provides repulsive force. The DGP brane acts as a conducting
plane in electrostatics (keeping in mind that in gravity different charges repel).
The repulsive nature of the brane requires a certain localization mechanism.
When the particle overcomes the localizing potential it rapidly moves to
the bulk. Particles of mass $m>8\pi^2M_*$ form a black hole within $1/M_*$ distance
from the brane.

\end{abstract}

\newpage

\section{Introduction}
\setcounter{equation}{0}

A phenomenologically acceptable 
five dimensional brane world theory with one infinite extra dimension 
was recently developed in \cite{dgp,dgkn,SG}. 
The low scale of quantum gravity $M_*$ is pulled
(renormalized) at the brane by the high scale $M_{SM}$ that describes the brane
localized standard model. More precisely, the 4D Einstein Hilbert term, 
with strength $\sim M_{\rm P}^2\sim M_{SM}^{2}$, is induced on the brane.
This effect ensures that the observer on the brane
sees weak 4D gravity (Newton constant $G_{N}\sim 1/M_{\rm P}^2$) 
up to the distance $r_c=M_{\rm P}^2/ M_{*}^3$. At distances bigger
than $r_c$ gravity becomes five dimensional. At short distances
gravity is modified by quantum corrections at $M_{*}^{-1}$.
Short distance gravity measurements exclude modification of laws
of gravity at distances bigger than $\sim$0.1mm$\sim 1/10^{-3}$eV \cite{measurements}.
Cosmological observations on the other hand suggest that gravity is
not changed to distances of order $\sim 10^{29}$mm. 
Thus, present knowledge about
gravity constrains the scale of gravity in this class of models to the range

\beq\label{constr}
10^{-3}\mbox{eV}<M_{*}<10\mbox{MeV}.
\eeq

Relativistic corrections, and the question how are they encoded
in the tensor structure of the graviton propagator were studied
in \cite{dis}. The cosmological consequences of the model and
especially the fact that the model gives rise to the 
accelerated universe (as observed \cite{rei})
were considered in \cite{cos}. The relevance for the solution to the cosmological
constant problem was considered in \cite{cc}

In the present paper we study how the induced 4D Einstein Hilbert term
affects dynamics of the particle of mass $m$ at distance $y_0$ from
the brane. This question is important in order  to identify
experimental signatures of collider black hole production in
this class of models. 
Since the particle itself induces the metric, there is
no static metric in which one could study the behavior of geodesic lines 
(that describe the motion of test particle). Instead of trying
to solve the problem in full relativistic theory, we will limit ourself
to Newtonian approximation and use some basic facts about black holes.
The result that we find is that the brane
repels particles into the bulk where they have lower (bigger in magnitude
and negative) self energy.

In the next section we derive and briefly discuss Newtonian potential.
In the third section we find the dependence of self energy of a particle
as a function of
its distance from the brane. Once particle leaves the brane, the
gradient of self energy forces it
to go from the brane to the bulk. We also describe the process of black hole
formation for particles with mass $m>8\pi^2M_*$. In the fourth section we present
the 'electrostatic' analog derivation of repulsive force for particles with 
mass $m<8\pi^2M_*$. Finally, in discussion, we address questions regarding
phenomenological consequences of the repulsive nature of the brane.

\section{Newtonian potential} 
\setcounter{equation}{0}

Action for the model \cite{dgp} is the sum of the 5D Einstein Hilbert term
and induced the 4D term on the brane

\beq\label{act}
S~=~ M_*^3\int d^4xdy \sqrt {G}{\cal R}_{(5)}+
 M_{\rm Pl}^2\int d^4x\sqrt{|g|}R.
\eeq
Here we divide 5D coordinates into 4D part (Greek indices) and extra coordinate
$y$ like $X^{A}=(x^{\mu},y)$, $G_{AB}$ is 5D metric and ${\cal R}_{(5)}$
its curvature scalar, $g_{\mu\nu}(x^{\mu})=G_{AB}(x^{\mu},y=0)
\delta_{\mu}^{A}\delta_{\nu}^{B}$ is induced 4D metric (we take straight
brane) and $R$ corresponding scalar curvature. The tension of the brane is taken
to be zero. If we take the limit of slowly varying weak fields, equations
of motion reduce to the equations for deviation of the $g_{00}$
component from flat space constant value (scalar gravity). The equation for
(Euclidean) Green's function
for the scalar gravity case reads 
\beq\label{gfp}
\left(\Box_4(1+r_{c}\delta(y))-\partial_{y}^2\right)G(x-x_0,y,y_0)=
\delta^{4}(x-x_0)\delta(y-y_0),
\eeq
which has the solution (see  \cite{SG})
\beq\label{gfwa}
G(p,y,y_0)=\frac{1}{p}e^{-p|y-y_0|}-\frac{1}{p}e^{-p(|y|+|y_0|)}
\frac{1}{1+1/r_cp}.
\eeq 
Let us evaluate the exact Newtonian potential at the point
$(\vec x, y)$ due to a static source of mass $m$ located at the position
$(\vec x', y_0)$. The potential is given as a Fourier transform 
of the Green's function (\ref{gfwa}) integrated over the time 
\beqa\label{newp}
V(r,y,y_0)= -\frac{m}{16\pi^2 M_{*}^3}  \Biggl(
\frac{1}{r^2+(y-y_0)^2}-\frac{1}{r^2+(|y|+|y_0|)^2} \nn \\ 
-\frac{i}{2rr_c}e^{ (|y|+|y_0|-ir)/r_c }  \left( \Gamma_0((|y|+|y_0|-ir)/r_c)-
e^{2ir/r_c}\Gamma_0((|y|+|y_0|+ir)/r_c)\right) \Biggr).
\eeqa 
Here $r=|\vec x -\vec x'|$ and $\Gamma_0(z)$ is an incomplete gamma function
(see appendix). The potential (\ref{newp}) can be expanded in powers of
$1/r_c$
\beq\label{prc}
V^1=-\frac{m}{16\pi^2 M_{*}^3}\frac{1}{rr_c} \arctan\frac{r}{|y|+|y_0|}
\eeq
\beq\label{drc}
V^2=-\frac{m}{16\pi^2 M_{*}^3}\frac{1}{rr_c}\left(
 \frac{(|y|+|y_0|)}{r_{c}}\arctan\frac{r}{|y|+|y_0|}+
\frac{1}{2}\left(\frac{r}{r_{c}}\right) \ln\frac{r^2+(|y|+|y_0|)^2
}{r_{c}^2}-(1-\gamma)\left(\frac{r}{r_{c}}\right) \right),
\eeq
where $\gamma\approx 0.5772$ is Euler's constant and superscripts on
potential denote terms in expansion. The potential to  first 
order in $1/r_c$ was discussed in detail in \cite{SG}.
Let us briefly discuss potential (\ref{newp}).
If the mass is on the brane ($y_0=0$), the first two terms in (\ref{newp})
cancel. The potential on the brane $(r,y=0)$, at distances 
$r\ll r_c$,  is four dimensional and the Newton's constant is
 $~G=1/(32\pi M_{*}^3r_c)$. 
As $~r/r_c~$ increases towards one, the second term in
(\ref{drc}) weaken its strength. Finally, for $r\gg r_c$ ($pr_c\ll 1$)
the potential becomes
purely five dimensional (the first term in equations (\ref{gfwa}),(\ref{newp})).
Similar behavior occurs if one looks at the potential at $(r=0,y)$. For small
$y$ it is a weak four dimensional potential with constant $2G/\pi $.
For $y/r_c\gg 1$ one can find the form of the potential by expanding
(\ref{newp}) (see appendix) and again obtain the expected five dimensional behavior.
 Up to a constant, potentials have the following short distance expansion
and asymptotic behavior

\beq\label{pr1} 
V(y=0,r\ll r_c)=-\frac{m}{32\pi M_{*}^3}\frac{1}{rr_c}\left(1+\frac{2}{\pi}
\frac{r}{r_c}\ln \frac{r}{r_c} \right),\quad
V(y=0,r\gg r_c)=-\frac{m}{16\pi^2 M_{*}^3r^2},
\eeq
\beq\label{pr2} 
V(r=0,y\ll r_c)=-\frac{m}{16\pi^2 M_{*}^3}\frac{1}{|y|r_c}\left(1+
\frac{y}{r_c}\ln \frac{y}{r_c} \right),\quad
V(r=0,y\gg r_c)=-\frac{m}{16\pi^2 M_{*}^3|y|^2}.
\eeq
Similar expansions can be easily obtained for a potential at any 'angle' in
the $r-y$ plane.

If the mass is in the bulk we have two different cases. For particles on
opposite sides of the brane, the first two terms in (\ref{newp}) cancel and
particles interact via the weak four dimensional gravity at distances
$\sqrt{(|y|+|y_0|)^2+r^2}\ll r_c$. That means that the brane is shielding
one side of the brane from the five dimensional gravitation of sources
on the other side of the brane. The effective radius of shielding is 
$\sim r_c$. If the sources are on  the same side of the
brane, the interaction is dominated by the first two terms  in (\ref{newp}).
Masses, sufficiently far from brane, basically interact via
strong five dimensional gravity. 

One can illustrate this behavior by plotting
the contours of the constant potential of the body as it moves from
the bulk towards the brane (Fig.1).
 At Schwarzschild radius $g_{00}\approx (1+2V)$ diverges.
 Although we don't have a relativistic solution to the system, one would expect that
the Schwarzschild surfaces (black hole horizons) behave as surfaces
of constant potential $V\approx -1/2$.

\vspace{5mm}
\centerline{\epsfig{file=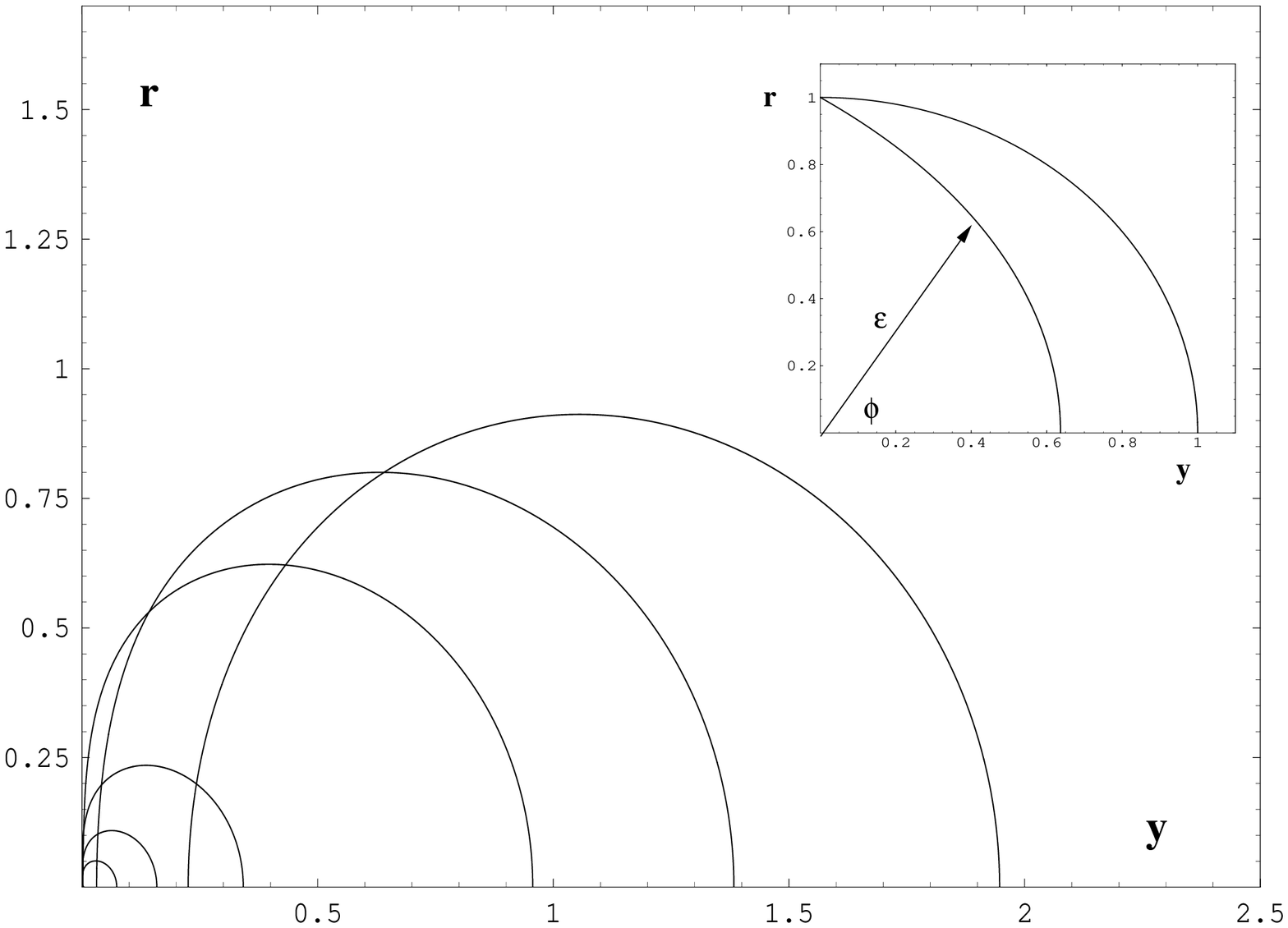,width=12cm}}
{\footnotesize\textbf{Figure 1:} Surfaces of constant
potential $V=-(1/16\pi^2)m/M_*$. Point mass is
on $r=0$, $y_0=1,0.5,0.2,10^{-2},10^{-3},10^{-4}$ in units of $M_{*}^{-1}$.
Inset: anisotropic cutoff distance $\epsilon$ for a mass on the brane}
\vspace{5mm}

\section{Self energy}
\setcounter{equation}{0}

In this section we will evaluate the classical self energy of the 
particle of mass $m$ in the presence of the DGP brane. Our main assumption, along the
lines of \cite{SG}, is that the gravity is cut off at distances $\epsilon\sim 1/M_*$.
Classically, gravitational self energy is
determined by the cutoff distance and the form of the potential.
Since the potential changes with the position, the gravitational  self 
energy of a particle will  be a function of its distance from the brane.
Gradient of self energy will give rise to a force that will try to 
move particle to the region of lowest gravitational self energy.
For $m<8\pi^2M_*$ we will use Newtonian approximation, since the potential
is weak at the cutoff distance $\sim M_{*}^{-1}$. For $m>8\pi^2M_*$,
we will use a Newton-like approximation, which will incorporate some basic
facts of general relativity.

\subsection{Case $m<8\pi^2M_*$}

For $m<8\pi^2M_*$ Schwarzschild radius in the bulk $r_{S5}= (1/M_*)\sqrt{m/8\pi^2M_*}$
\cite{myers} is  smaller than the
cutoff $M_{*}^{-1}$ so Newtonian approximation is
justified. Let us remind how we calculate gravitational self energy of a particle. One
wants to find an energy gained by assembling
a particle of mass $m$. Equivalently, one can find the energy needed
to destroy a particle by taking away infitesimal pieces of matter and removing
them to infinity, where the potential is defined to be zero. If the force
starts to act at a distance $\epsilon$ from the center of mass distribution of
a particle, one then finds the expression for self energy to be

\beq\label{seta}
W=-\int_{0}^1(1-M)dM\int_{\epsilon}^{\infty} \frac{dV(r)}{dr}dr=
\frac{1}{2}V(\epsilon).
\eeq
In our normalization of Newton constant, that leads to self energies 
in four dimensional theory (on the brane) and five dimensional
theory (infinitely far from the brane)
\beq\label{pfft}
W_4=-\frac{1}{64\pi}\left(\frac{m}{M_{\rm P}}\right)^2\frac{1}{\epsilon},\quad
W_5=-\frac{1}{32\pi^2}\left(\frac{m^2}{M_{*}^3}\right)\frac{1}{\epsilon^2}.
\eeq
If we take that the gravity cutoff $\epsilon$ is just an inverse scale of gravity
$M_*$, then the ratio of self energies in pure 4D and 5D theories is
\beq\label{rati}
\frac{W_4}{W_5}=\frac{\pi}{2}\left(\frac{M_*}{M_{\rm P}}\right)^2.
\eeq
In  pure four or five dimensional theories energy of the mass density
$\rho(x)$ is given by $W=(1/2)\int\rho(x)V(x)d^nx$. One can then use Gauss
law to relate mass density and divergence of the field. After partial
integration energy can be written as an integral over the square of the field
$W\sim -\int|\nabla V(x)|^2d^nx$. By integrating energy stored in the field
in pure 4D and 5D theories one again finds expressions (\ref{pfft}). We
must stress here that the discussed theory  is neither
purely 4D or 5D theory. In particular, for the mass
 on the brane, Gauss law is not valid 
(if we call $r_5$ the 5D radial distance from the mass at $y_0=0$, then
the field drops like $\sim 1/r_{5}^2$, while the surface area increases like
$\sim r_{5}^3$). For this reason we will use (\ref{seta}) when calculating 
self energy of the mass at an arbitrary position in 5D space (one should not use
Gauss' law).

Let us look at the self energy of the particle on the brane. By using
prescription (\ref{seta}) we find that the self energy 
is that of a four dimensional particle. However, since our space is not isotropic
there is an apparent ambiguity in self energy. It depends on the direction
from which  we assembled particle.  Let us define polar coordinates
$\rho=r^2+y^2$ and $\phi=\arctan(r/y)$ and assemble particle at $y_0=0,r=0$
by bringing infitesimal masses from infinity and direction $\phi$ from the bulk .
Self energy of the particle on the brane then varies by a factor $\pi/2$
(same as Newton constant) for angles $0<\phi<\pi/2$:

\beq\label{ambi}
W(y_0=0)=-\frac{1}{32\pi^2}\left(\frac{m}{M_{\rm P}}\right)^2
\frac{\phi}{\sin\phi}\frac{1}{\epsilon}.
\eeq
Since the self energy must be a well defined quantity, we conclude
that the factor $\phi/\sin\phi$ defines the physical cutoff distance 
when the mass is on the brane. Certainly, the space in question is not
isotropic, and we cannot assume that the gravity cutoff surface
is a 3-sphere, but rather a surface of constant field strength with  
average distance from the particle $\sim 1/M_*$ (inset to Fig.1).
For the arbitrary position of particle $y_0$,  the expression for
self energy to first order in $1/r_c$ is

\beq\label{seap} 
W(y_0)=-\frac{1}{32\pi^2}\left(\frac{m^2}{M_{*}^3}\right)
\left(    \frac{1}{\epsilon^2}\left(1-\frac{1}{1+4(y_{0}/\epsilon)^2
+4(y_0/\epsilon)\cos\phi}\right)
      +\frac{1}{r_c\epsilon}\frac{1}{\sin\phi}\arctan\left( \frac{\epsilon\sin\phi}
{2y_0+\epsilon\cos\phi}\right) \right).
\eeq
For $y_0>\epsilon$, dependence on $\phi$ can be neglected and the self energy
has the form
\beq\label{phneg}
W(y_0)=W_5\left(1-\frac{1}{1+4(y_0/\epsilon)^2}\right)+W_4\frac{1}{\pi(y_0/\epsilon)}.
\eeq
For $0<y_0<\epsilon$ (neglecting the small $1/r_c$ contribution of 4D self energy)
, one gets the same answer by following argument.
The first term in (\ref{seap}) represents isotropic 5D interaction and should
be cut off at the surface of 3 sphere of radius $\epsilon$. The second term
is anisotropic contribution and its cutoff
 has to be defined so that it doesn't depend on
the angle $\phi$. If we define (angle) dependent cutoff for the second term
as $\tilde{\epsilon}^2=\epsilon^2f(y_0/\epsilon,\phi)$ one finds that 
$f(y_0/\epsilon,\phi)=(-2(y_0/\epsilon)\cos\phi +\sqrt{1+4(y_{0}/\epsilon)
^{2}\cos^{2}\phi})$. Plugging this back to (\ref{seap}) one obtains the behavior
of 5D contribution as in (\ref{phneg}).

To summarize, at the brane, the particle has 4D self energy, upon leaving the
brane, within a couple of $\epsilon$ distance, it gains the biggest part of its 5D self
energy and looses its 4D self energy. The particle at $y_0>\epsilon$ will feel strong 
force

\beq\label{fors}
F_y(y_0)=-\frac{dW(y_0)}{dy_0}=\frac{1}{64\pi^2}\frac{m^2}{M_{*}^3}\frac{1}{y_{0}^3}
\left(1-\frac{y_0}{r_c}\right).
\eeq
This force will try to push the particle
 to the bulk where its self energy increases in magnitude
by the large factor of $(M_{\rm P}/M_*)^2$.

\subsection{Case $m>8\pi^2M_*$}

If the mass of the particle is bigger than the scale of gravity $M_*$,
we cannot calculate the self energy by cutting off the Newtonian potential at
$\epsilon\sim M_{*}^{-1}$. The reason is that Schwarzschild radius in
the bulk is bigger than the inverse scale $M_*$, and at distances shorter 
than Schwarzschild radius $r_{S5}$  gravity is not weak. In following 
considerations we will not write negligible corrections of the order of $1/r_c$.

Let us calculate the self energy of a 5D black hole by approximating the black hole
as an object that gravitates via Newton's law at distances bigger than
event horizon $r>r_{S5}$. What happens with potential at distances below
event horizon doesn't influence the energy of the world outside horizon. 
 Again we construct the self energy by assembling black hole at the origin out of 
infitezimal pieces of matter located at infinity. To assemble the point mass
$M_*$ we need to bring matter to a cutoff distance $M_{*}^{-1}$, since the 
Schwarzschild radius is smaller then the inverse cutoff. This contribution  equals to
$W_5(m=M_*)=-M_*/(32\pi^2)$.
Work gained in bringing the rest
of the mass (from $M_*$ to $m$) would be the work done in bringing it 
only to the Schwarzschild radius. Outside the horizon there is no change in
energy no matter how the potential changes below the horizon. 
Now we introduce a loose definition of the Schwarzschild radius as a radius
at which Newtonian potential has the value $-1/2$. Since the potential
energy gained by bringing the mass $dM$ from infinity
to Schwarzschild radius is by (our) definition $dM/2$, the total self energy
of a 5D black hole of mass $m$ is

\beq\label{5dbh}
W(m>M_*,y_{0} \longrightarrow \infty)=W_5(m=M_*)-\frac{1}{2}\int_{M_*}^{m}dM=
-\frac{1}{32\pi^2}M_*-\frac{1}{2}\left(m-M_*\right).
\eeq
In our simplified model of a black hole, self energy is negative and 
of the order of (factor of $1/2$) the mass of a black hole. It is interesting
to note that 
self energy could, in principle, be equal to the rest mass so that it wouldn't
cost anything to produce it. The situation is reminiscent of the fact that
the total mass of universe, Newton's constant and the Hubble radius conspire
in such a way that it might not cost anything to create particles at the center
of the universe, since
their rest mass energy is of the order of their gravitational (negative) energy.

Let us see how the formation of the black hole happens as we remove a particle
of mass $m>8\pi^2M_*$ from the brane.
On the brane, the Schwarzschild radius  $r_{S4}\sim M_{\rm Pl}^{-1}(m/M_{\rm Pl})$
is much smaller than the cutoff distance $M_{*}^{-1}$. For that reason the self
energy on the brane is given by $W_4$ (\ref{pfft}). The self energy on the
brane  is much smaller 
than the self energy far away from the brane (\ref{5dbh}) and can be neglected.
Thus, particles of mass $m>8\pi^2M_*$
 will (as well as particles with $m<8\pi^{2}M_*$) 
be repelled from the brane to the bulk where their self energy is lower.
However, the character of the repelling force will differ from the case of
$m<8\pi^2M_*$. 
Let us define the Schwarzschild surface as a surface on which $V(r,y,y_0)=-1/2$.
When the particle is removed from the brane, the Schwarzschild surface expands
from a point (actually, the three sphere of radius $r_{S4}\sim M_{\rm Pl}
^{-1}(m/M_{\rm Pl})$ ) anisotropically (Fig.1).
After the particle reaches
a certain value of $y_0$, the Schwarzschild surface crosses the three sphere of radius
$M_{*}^{-1}$ that describes the gravity cutoff radius (Fig.2) . This
crossing first happens for
the value of  $\phi=0$. Up to that point the self energy and the force
on particle are the same as in the case $m<8\pi^2M_*$.
After that point we cannot consider
$M_{*}$ self energy cutoff distance and the self energy evolves different
from the case $m<8\pi^2M_*$. Moving the particle further into bulk, the Schwarzschild
surface grows and takes over the $M_*$ sphere at larger angles and
 finally,  for some critical
value of $y_0$, the $M_*$ sphere becomes completely contained inside
the Schwarzschild surface. At that point we can say that the black hole 
formation is finished and the particle will have (up to corrections of the
 order of $M_*$) the self energy of a 5D black hole.

\vspace{5mm}
\centerline{\epsfig{file=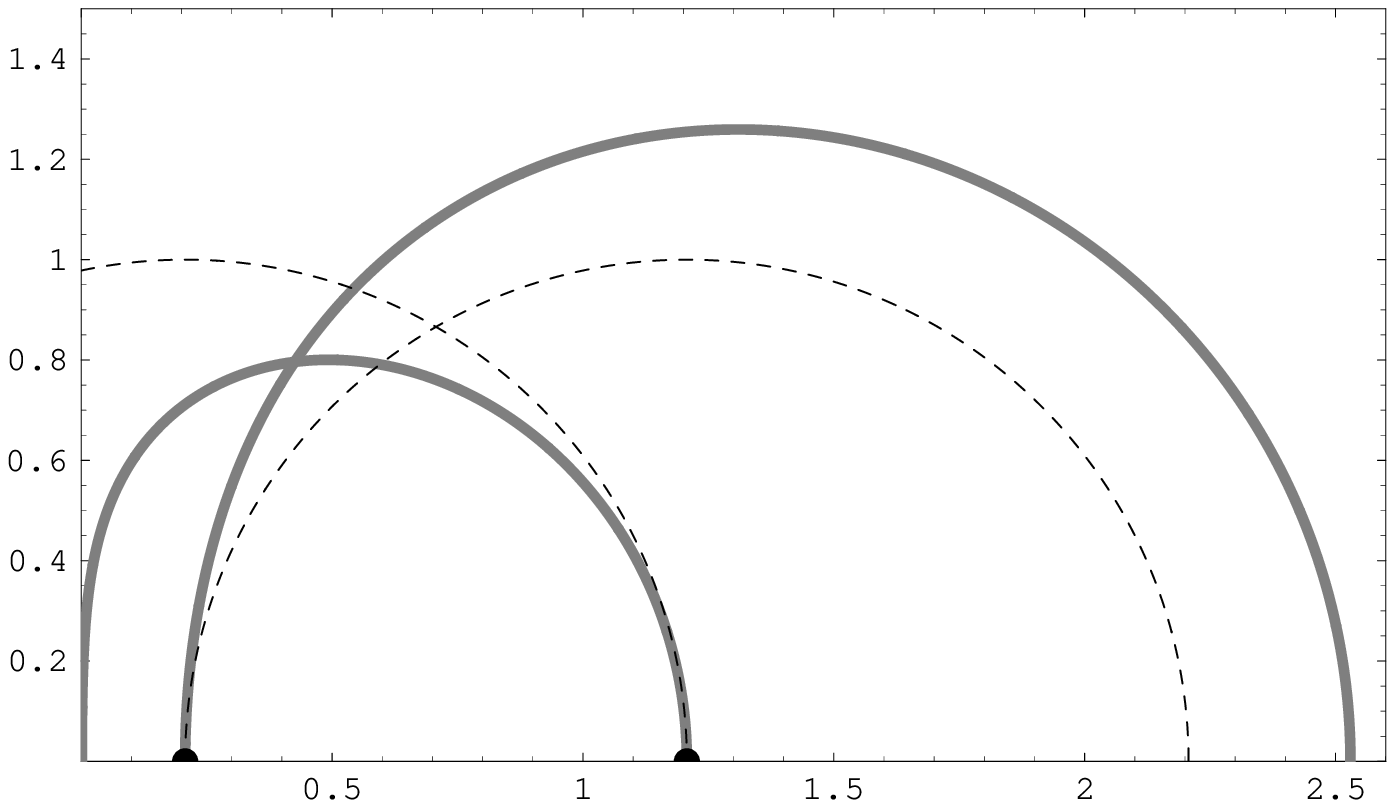,width=12.5cm}}
{\footnotesize\textbf{Figure 2:}   The figure shows the formation of the black hole
horizon in the $r-y$ plane. The horizontal axis measures distance from the brane $y$
and the vertical axis measures the distance along the brane $r$, 
both in units of $M_{*}^{-1}$. The mass of the
particle is taken to be $m=16\pi^2M_*$ i.e. twice the critical mass. Solid lines
represent Schwarzschild surfaces, dashed lines represent surfaces of three
spheres of unit radius and masses are represented by dots. 
The horizon emerges when the particle is at $y_{0}^{\phi=0}=0.207$ 
and completely encloses the $M_*$ sphere one unit farther.   }
\vspace{5mm}

Points where the Schwarzschild surface crosses the $M_*$ sphere for $\phi=0,\pi$
can be expressed from
potential (\ref{newp})

\beq\label{crosp}
y_{0}^{\phi=0}=\frac{1}{2M_*}\left( \frac{1}{\sqrt{1-8\pi^2M_*/m}}-1  \right),
\quad y_{0}^{\phi=\pi}=y_{0}^{\phi=0}+\frac{1}{M_*}  .
\eeq
For $m\gg 8\pi^2M_*$ these expressions become $y_{0}^{\phi=0}\approx 2\pi^2/m$ and
 $y_{0}^{\phi=\pi}\approx 1/M_*$. Thus, the horizon starts
forming at $\sim 2\pi^2/m$ and
is formed precisely $1/M_*$ farther. The force felt by the particle on $y_0<2\pi^2/m$
is the same as in the case $m<8\pi^2M_*$ ($y_0 M_{*}\ll 1$)

\beq\label{bhapf}
F_{y}(y_0<2\pi^2/m)=\frac{1}{4\pi^2}m^2M_*y_0.
\eeq
For $2\pi^2/m<y_0<1/M_*$, neglecting terms of order $M_*$ in self energy, force is
approximately

\beq\label{bhapf1}
F_{y}(y_0>2\pi^2/m)\approx \frac{\pi^2}{y_{0}^2}.
\eeq
To summarize, for $m>8\pi^2M_*$, the 5D Schwarzschild radius is bigger than
the cutoff $M_{*}^{-1}$, and one cannot use the Newtonian theory to calculate 
the self energy of the particle. Modeling a black hole as an object that 
gravitates with Newtonian potential outside the horizon, we calculated self energy
and estimated the force felt by the particle. As in the case $m<8\pi^2M_*$
particles are repelled from the brane.

\section{Conductor analogy}
\setcounter{equation}{0}

In this section we will rederive results of the previous section for particles
with $m<8\pi^2M_*$ from a
different point of view. The lagrangian of our theory can be  thought of
as the lagrangian for a purely 5D theory with specific type of source localized
at $y=0$. One would expect this correspondence to be valid everywhere
except at $y=0$ (worldvolume of the source) because the source 
itself is a kinetic term for the 4D theory. At $y=0$, the value of delta function
diverges and the 4D kinetic term becomes dominant.
Our  gravity theory in this approach $(y\neq 0)$ becomes equivalent to
a 5D gravity  in the presence of an infinite 3-plane with a 
specific mass (charge) distribution.  
In the Newtonian approximation theory is equivalent to the electrostatic setup
of a charge near the conducting plane.
We will use  symbol $\vec E$ for the gravitational
field and sometimes interchange the terms mass and charge.

Let us take mass $m$ at position $(\vec r=0, y_0)$ and look at the
field at position $(\vec r, y)$. One can ask what kind of charge distribution
on the plane would produce potential (\ref{newp}).

\vspace{5mm}
\centerline{\epsfig{file=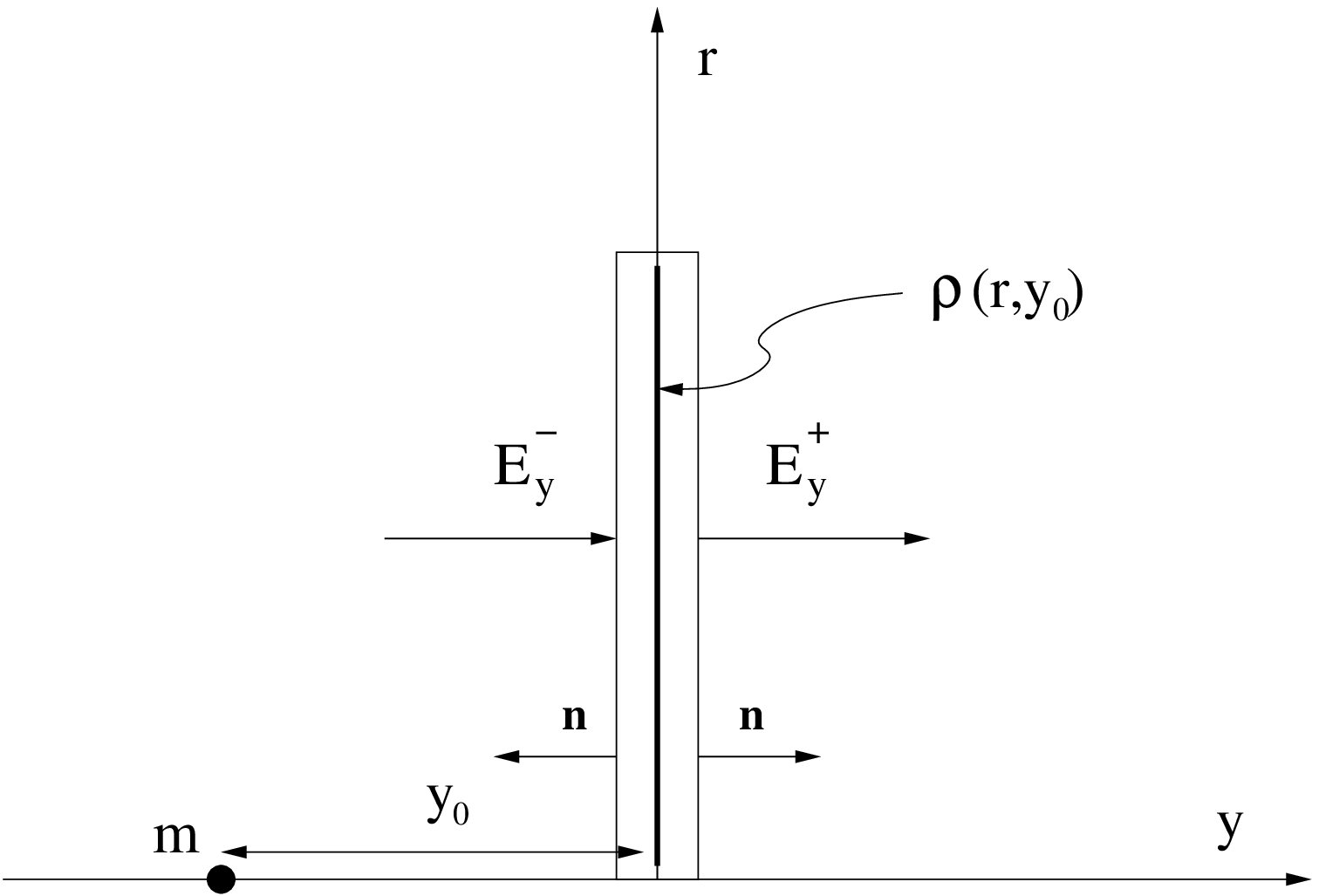,width=11cm}}
{\footnotesize\textbf{Figure 3:} Gaussian pillbox for determination
of the effective charge distribution $\rho(r,y_0)$ induced by the charge $m$ at
position $(r=0,y_0)$ }
\vspace{5mm}

A component of the field in the $y$ direction is discontinuous at $y=0$ with
discontinuity (to a first order in $1/r_c$)

\beq\label{disco}
(E_{y}^{y=+0}-E_{y}^{y=-0})= \frac{m} {4\pi^2M_{*}^3} \frac{|y_0|-(r^2+y_{0}
^2)/(2r_c) }{ (r^2+y_{0}^2)^2}
\eeq
Applying the Gauss theorem on the 5D pillbox, as shown in Fig.3, we can
find the charge distribution on the plane

\beq\label{gauss}
\nabla\cdot {\bf E}=-\frac{1}{4M_{*}^3}\rho(\vec{x},y)
\quad \longrightarrow \quad \rho(r,y_0)=
-\frac{1}{\pi^2}\frac{m|y_0|}{(r^2+y_{0}^2)^2}+\frac{m}{2\pi^2r_c(r^2+y_{0}^2)}.
\eeq
The first term in the expression for charge density (\ref{gauss}) represents 
the distribution of negative charge, sharply peaked around $r=0$, i.e. the projection
of the charge position on the brane. Its integral over the volume of the
brane is independent of $y_0$ and equals precisely $-m$. As $y_0\longrightarrow 0$
this term approaches distribution of a point-like charge $-m\delta(\vec r)$.
The second term in (\ref{gauss}) is a small (notice $1/r_c$ suppression) distribution
of positive charge, much less localized then the negative charge distribution. 
In order to find the total induced charge density we should integrate
the exact expression obtained from the potential (\ref{newp}) that is correct 
to all orders in $1/r_c$. The integral of the charge distribution due to 
the third term in (\ref{newp}) can be obtained numerically and is equal to $m$.
Thus, the total charge induced on the brane due to the charge $m$ in the bulk
is zero, as one would expect.

Having the charge distribution, we can calculate the interaction energy
between the mass and the induced distribution
 ('image' distribution of mass $-m$ and the background
distribution of mass $+m$). For $y\ne 0$ our theory is just the 5D Newtonian gravity
so the potential energy of interaction is (to first order in $1/r_c$)

\beq\label{inten}
W(y_0)=\frac{1}{2}\int\rho(x)V(x)dV=-\frac{m}{32\pi^2M_{*}^3}
\int\frac{\rho(r,y_0)4\pi r^2dr}{r^2+y_{0}^2}=\frac{1}{128\pi^2}\left(
\frac{m^2}{M_{*}^3}\right)\frac{1}{y_{0}^2}\left(1-\frac{2y_0}{r_c}\right).
\eeq
Cutoff effects in this derivation were neglected, so it is understood 
that $y_0>\epsilon $. The result (\ref{inten}) coincides with (\ref{phneg})
and gives the same force (\ref{fors}).

Let us summarize what happens in our 5D analog picture. Charge $m$ in the
bulk induces negative 'image' charge distribution of total charge $-m$, localized
at the $r=0$, and the small uniform background positive mass distribution.
The total induced charge on the brane is zero. As charge $m$ approaches the brane
$(y_0\longrightarrow 0)$ image charge distributions tends to the distribution of a
 point like charge $-m$, which strongly repels mass $m$. Finally, charge $m$ and
the image $-m$ annihilate and $m$ distributes itself uniformly on the brane.
Described process is
completely analogous to the behavior of the charge near the conducting plane.
The only difference is that in electrostatics charges of the opposite sign attract
and in gravity they repel. Thus, mass $m$ in the bulk is repelled from
the brane by its image $-m$. In this sense the DGP brane acts as a gravity conductor,
shielding the fields and giving rise to a repulsive force.
One could imagine constructing tensionless objects with this property that would 
gravitationally repel
masses or act as gravitational dipoles. In cosmological setups the tensionless DGP 
brane would gravitationally shield ('shadow') parts of the universe and could 
modify cosmological evolution.

\section{Discussion}
\setcounter{equation}{0}

In previous sections we showed that the particles are repelled from
the brane that induces kinetic term. By analogy with an ordinary wall with a tension
\cite{Sikivie}, we can loosely say that the induced kinetic term
creates localized energy momentum density on the brain in which repulsive
tension dominates over attractive energy density. 
Theories with low scale of gravity predict collider production of black holes.
Because of the repulsive nature of the brane, black holes produced in
collider experiments would be repelled to the bulk. 

The repulsive nature of the brane requires certain localization mechanism
for standard model particles. We can distinguish two different cases. 
In the first case standard model particles are entities that cannot  exist
independent of the brane. Well known examples are goldstone modes of
broken translational invariance (elastic waves of the brane), modes
of open strings with endpoints stuck on the brane, or simply fermionic
zero modes on the soliton-like wall. In this case, 
particles feel force but they cannot escape to the bulk.
Other possibility is that the standard model particles
are entities that exist independent
of the brane. Then, we have to introduce localizing potential $\Delta W$ that
keeps them on the brane.
Since on colliders we don't see events in which particles
just disappear, the depth of the localizing potential $\Delta W$ would have to
be bigger than $\sim 1$Tev. From the present bound on the size
of universal extra dimensions one knows that the range of the
localizing potential should be less than 300Gev \cite{apelq}.
Localizing potential can be due to short range (contact) interactions with
the matter of the first type, or the brane itself. For phenomenologically
acceptable energy densities on the brane, the gravitational attraction
can not provide localizing potential.
The particle localized on the brane will feel effective
potential which is a combination of short distance localizing potential and
repulsive potential. With the potential of depth $\Delta W$, all particles
lighter than $\Delta W$ will be in stable equilibrium on the brane.
Particles heavier than $\Delta W$ would be in a metastable state
on the brane, because their self energy in the bulk is roughly their mass
(\ref{5dbh}). Metastable particles can then tunnel through the barrier
into the bulk.
The brane can, in principle, be populated with both types of particles. Intrinsically
brane particles would be stable (with respect to escape to the bulk decay).
Particles trapped on the brane, on the other hand, can decay
by escape to the bulk. Upon leaving the 
brane those particles would gain energy of the order of their mass (bulk self energy)
in the vicinity of the brane (distance $\sim 1/M_*$). The recoil effect of the brane
would produce stable particles (goldstone modes, zero modes) with a total
energy of the order of the mass of particle that escaped to the bulk. This kind of
decay to the bulk would make missing energy signal on colliders smaller 
than one expected in a scenario with an ordinary brane.

So far, all the discussion has referred to a brane
of infitezimal thickness (delta function
type brane). Real, physical branes, have finite thickness. It would
be interesting to see how the finite thickness
affects particle dynamics and if it can provide localization mechanism.
 Arguments that we used in the derivation
of the repulsive force in 5D model, apply equally well to branes in space with more
than one extra dimension. To completely understand particle
dynamics, one would certainly like to have an exact relativistic solution.

\vspace{1cm}
{\bf Acknowledgments}
\vspace{0.1cm} \\

I would like to thank Georgi Dvali, Andrei Gruzinov,
Arthur Lue, Gregory Gabadadze and Engelbert Schucking 
for helpful discussion and reading the
manuscript. This work was supported in part by David
and Lucille Packard Foundation, grant number 99-1462.

\vspace{0.2in}
\newpage

\begin{appendix}

\section{incomplete gamma function}


Incomplete gamma function $\Gamma_{\alpha}(z)$ is defined as

\beq\label{incgam}
\Gamma_{\alpha}(z)=\int_{z}^{\infty}e^{-t} t^{\alpha -1}dt.
\eeq

It satisfies

\beq\label{relg}
\frac{d}{dz}\Gamma_0(z)=-\frac{e^{-z}}{z},\quad\quad 
\int \Gamma_0(z)dz=-e^{-z}+z\Gamma_0(z).
\eeq

For small and large values of argument it has following expansions

\beqa\label{expas}
\Gamma_0(z)\approx -\gamma -\ln z +z+{\cal O}(z^2),\quad z\longrightarrow 0\nn \\
\Gamma_0(z)\approx e^{-z}\left( 1/z-1/z^2+2/z^3+{\cal O}(1/z^4)\right),
\quad z\longrightarrow \infty.
\eeqa

\end{appendix}

\end{document}